\documentstyle[11pt,aaspp4]{article}
\input psfig.tex

\begin{document}

\title{On the energy of gamma-ray bursts}
\author{Deborah L. Freedman$^{1,2}$ \& Eli Waxman$^1$}
\affil{$^1$Department of Condensed-Matter Physics, Weizmann Institute, Rehovot
   76100, Israel}
\affil{$^2$Princeton University Observatory, Princeton, NJ 08544, USA}

\begin{abstract}

We show that $\gamma$-ray burst (GRB) afterglow observations strongly suggest,
within the fireball model framework, that radiating electrons are shock
accelerated to a power-law energy distribution,
$dn_e/d\gamma_e\propto\gamma_e^{-p}$, with
universal index $p\approx2.2$, and that the fraction of shock energy
carried by
electrons, $\xi_e$, is universal and close to equipartition, $\xi_e\sim1/3$.
For universal $p$ and $\xi_e$, a single measurement of the
X-ray afterglow flux on the time scale of a day
provides a robust estimate of the
fireball energy per unit solid angle, $\varepsilon$, averaged
over a conical section of the fireball of opening angle $\theta\sim0.1$.

Applying our analysis to BeppoSAX afterglow data we find that:
(i) Fireball energies are in the range of $4\pi\varepsilon=10^{51.5}$ to
$10^{53.5}$~erg;
(ii) The ratio of observed $\gamma$-ray to total fireball energy
per unit solid angle, $\varepsilon_\gamma/\varepsilon$,
is of order unity, satisfying
$|\log_{10}(\varepsilon_\gamma/\varepsilon)|\lesssim0.5$;
(iii) If fireballs are jet like,
their opening angle should satisfy $\theta\gtrsim0.1$.
Our results imply that if typical opening angles are $\theta\sim0.1$, a value
consistent with our analysis,
the total energy associated with a GRB event
is in the range of $10^{50}$~erg to $10^{51.5}$~erg.

\end{abstract}


\section{Introduction}

The widely accepted interpretation of the
phenomenology of $\gamma$-ray bursts (GRBs) is that the
observable effects are due to the dissipation of the kinetic energy
of a relativistically expanding fireball
(see \cite{Meszaros95} and \cite{Piran96} for reviews).
In the last two years,
afterglows of GRBs have been discovered in X-ray (\cite{Costa97}),
optical (\cite{Paradijs}) and radio (\cite{Frail97}) wavelength.
Afterglow observations lead to the confirmation of the cosmological origin
of the bursts through the detection of redshifted optical absorption lines
(e.g. \cite{Metzger97}), and confirmed (\cite{Waxman97a,Wijers97})
standard model predictions (\cite{Rhoads93,Katz94,MnR97,Vietri97a})
of afterglow
that results from the collision of the expanding fireball with
surrounding medium.

In fireball models of GRBs the energy released by an explosion
is converted to kinetic energy of a thin baryonic shell expanding at an
ultra-relativistic speed. After producing the GRB, the shell impacts on
surrounding gas, driving an ultra-relativistic shock into the ambient
medium. After a short transition phase, the expanding blast wave approaches
a self-similar behavior (\cite{BnM76}).
The long term afterglow is produced by the expanding shock that
propagates into the surrounding gas. This shock continuously heats
fresh gas and accelerates relativistic electrons to a power-law
energy distribution, which produces the
observed radiation through synchrotron emission. The simplest fireball
afterglow model therefore depends on five model parameters: the explosion
energy $E$, the ambient density $n$, the spectral index of the electron
energy distribution $p$, and the fractions $\xi_e$ and $\xi_B$ of shock
thermal energy carried by electrons and magnetic field respectively. The model
may be further elaborated by allowing inhomogeneous density distribution
and deviations from spherical symmetry
(\cite{MRW98,Vietri97b,Rhoads97,Chevalier99}).

Despite the fact that afterglow observations are in general consistent with
fireball model predictions, observations do not allow, in almost all
cases, the determination of
basic fireball parameters. In particular, the fireball
energy, and hence the efficiency with which this energy is converted to
$\gamma$-ray radiation,
can be reliably determined only in one
case, namely that of GRB 970508, for which wide spectral coverage is available
over hundreds of days (\cite{Waxman97b,WKF98,GPS99,WnG99,FWK99}).
For all other cases, energy estimates rely on the observed $\gamma$-ray
fluence (which dominates the fluence at all other wave-bands).
Such energy estimates are, however, subject to large uncertainties.

During $\gamma$-ray emission the fireball expands with a large Lorentz factor,
$\Gamma\sim10^{2.5}$, and a distant observer receives radiation from
a conical section of the fireball of opening angle $\sim10^{-2.5}$ around the
line of sight. Observed $\gamma$-ray fluence provides therefore the
$\gamma$-ray energy per unit solid angle $\varepsilon_\gamma$
emitted by a conical section of the fireball of opening angle $\sim10^{-2.5}$.
Thus, if the fireball is a jet of opening angle $\theta_j\ge10^{-2.5}$,
its total $\gamma$-ray energy would be smaller by a factor
$\theta_j^2/4$ compared to that inferred assuming spherical symmetry.
The total $\gamma$-ray energy emitted may differ
significantly from that obtained assuming spherical symmetry
also for $\theta_j\sim1$,
if $\varepsilon_\gamma$ is strongly dependent on the
angle $\theta$ with respect to the line of sight. It has
therefore been suggested (\cite{KnP99})
that the most energetic bursts
(i.e. those with highest isotropic $\gamma$-ray energy)
may not represent much higher energy release
from the source, but rather cases in which our line of sight happens to
coincide with a small patch on the fireball emitting surface which is much
brighter than average.

Energy estimates based on $\gamma$-ray fluence are
furthermore uncertain even
in the case of spherical symmetry, since it is possible
that only a small fraction of fireball energy is converted to $\gamma$-ray
radiation. It is generally argued, for example, that if $\gamma$-ray emission
is due to internal shocks within the fireball, then only a small
fraction, $<10^{-2}$, of fireball energy is converted to $\gamma$-rays
(\cite{Spada,Kumar,KnP99}). The low efficiency obtained in these analyses
is not due to low electron energy fraction (equipartition, $\xi_e\sim1/3$,
is assumed), but
rather due to the low efficiency of converting kinetic energy to thermal
energy in the models analyzed, and due to the fact that not all
of the radiation is emitted in the observed $\gamma$-ray band. For $\xi_e$
values below equipartition, the efficiency would be even lower.

The main goal of this paper is to address the open questions
associated with GRB energy and $\gamma$-ray production efficiency.
We show that significant progress can be made in inferring basic
fireball parameters if, instead of analyzing individual burst data,
the ensemble of GRB afterglow observations is analyzed under
the hypothesis that the values of parameters which are determined by
the micro-physics of the relativistic shock, e.g. $p$ and $\xi_e$,
are universal. We first show in \S2.1 that afterglow observations provide
strong support to the hypothesis that the spectral index of the
electron energy distribution is universal, $p\approx2$.
We then show in \S2.2 that, adopting a universal $p$ value,
a single X-ray afterglow flux measurement at time $t$ provides a
robust estimate of the total energy per unit solid
angle carried by fireball electrons, $\varepsilon_e\equiv\xi_e E/4\pi$,
over a conical section of the fireball of opening angle $\approx1/\Gamma(t)$,
where $\Gamma$ is the fireball expansion Lorentz factor
(For X-ray observations
on day time scale, $\Gamma\sim10$). We then show in \S2.3 that
afterglow observations also imply a universal value close to equipartition
for the electron energy fraction, $\xi_e\sim1/3$. Thus,
X-ray afterglow flux measurement also provides a robust estimate of the
total fireball energy per unit solid angle, $\varepsilon\equiv E/4\pi$.
Applying these ideas we provide in \S2 constraints on fireball energy
for all GRBs with X-ray afterglow data.
The implications of our results are discussed in \S3.

\section{Analysis}

\subsection{Electron spectral index}

For a power-law electron energy distribution,
$dn_e/d\gamma_e\propto\gamma_e^{-p}$, the spectrum of synchrotron emission
is $f_\nu\propto\nu^{-(p-1)/2}$ at frequencies where emission is
dominated by electrons for which the synchrotron cooling time is larger
than the source expansion time, and $f_\nu\propto\nu^{-p/2}$
at higher frequencies, where electron synchrotron cooling time is shorter
(e.g. \cite{Rybicki}). Unfortunately, observed afterglow spectra do not
in general allow an accurate determination of $p$, since optical spectra
may be
affected by extinction and since X-ray data typically does not allow
accurate determination of X-ray photon spectral indices. Accurate
determination
of $p$ is possible, nevertheless, in the case of GRB~970508, where
radio, optical and X-ray spectral data are available;
these data imply $p=2.2\pm0.1$
(\cite{Galama98,FWK99}), and in the
case of GRB~990510, where BVRI optical data determine
$p=2.2\pm0.2$ (\cite{Stanek99}).

Within the framework of the fireball model, the afterglow flux at high
frequency exhibits a power-law decay in time, $f_\nu\propto t^{-\alpha}$,
where $\alpha$ is related to the electron index $p$. Here too, accurate
determination of $p$ is generally not possible based on measurements of
$\alpha$, since the relation between $\alpha$ and $p$ depends on
the spatial distribution of ambient density and on the angular distribution
of fireball parameters (\cite{MRW98}). However, in several cases
the time dependence of optical flux strongly suggests a universal value
$p\approx2$. The steepening of optical flux decay in GRB~990510 afterglow
(\cite{Stanek99}) and the fast decline of optical flux for
GRB~980519, $\alpha=2.05\pm0.04$, and GRB~980326, $\alpha=2.1\pm0.1$,
(\cite{Halpern99,Groot99}) is most naturally explained
(\cite{Rhoads97},1999,\cite{MRW98,SPH99,Harrison99,Stanek99})
by the fireball being initially
a collimated jet. In this case steepening takes place once sideways
expansion of the jet becomes important, at which stage the
the temporal power-law flux decay index is $\alpha=p$. The observed values
of $\alpha$ are consistent therefore with $p\approx2$.

We present here additional evidence for a universal $p\approx2$ value.
In Table 1 the effective photon spectral index determined
by X-ray and optical afterglow fluxes, $\beta_{OX}\equiv-\ln(f_X/f_O)/
\ln(\nu_X/\nu_O)$, is shown
for all cases where both X-ray and optical afterglow
fluxes are available. The values of $\beta_{OX}$ are in the range of
0.6 to 1.1. This is the range expected for
a power-law electron energy distribution with $p=2.2$. For such a distribution
$\beta_{OX}=p/2=1.1$ is obtained for the case where the frequency $\nu_c$,
at which emission is dominated by electrons with synchrotron cooling
time comparable to the fireball expansion time, is below the optical
band $\nu_c\le\nu_O$, $\beta_{OX}=(p-1)/2=0.6$ is obtained for
$\nu_c\ge\nu_X$,
and $0.6<\beta_{OX}<1.1$ for $\nu_O<\nu_c<\nu_X$.

We note that an alternative model for the fast, $\alpha\approx2$, optical
flux decline of GRB~980519
 has been suggested (\cite{Chevalier99}), in which the afterglow
is produced by a spherical fireball expanding into a pre-burst massive
star wind, where $n\propto r^{-2}$. Since for $p\approx2$
expansion into a wind gives an optical flux decline which is not significantly
steeper than that obtained for expansion into homogeneous density
(e.g. \cite{LnW99}), the wind model for GRB~980519 invokes a steep
electron index $p=3$ to account for the fast decline (\cite{Chevalier99}).
We note, however, that for such a steep electron index the model X-ray
flux is $\sim6$ times lower than the measured flux at $t=0.5$~d\footnote{
The apparent agreement between model and measured X-ray flux in Fig. 1
of \cite{Chevalier99} is due to the fact that the presented flux is
calculated assuming $\nu_c>\nu_X$ while for the wind model parameters
$\nu_c\approx3\times10^{16}{\rm Hz}\ll\nu_X=10^{18}$~Hz.}.

\subsection{Fireball electron energy}

For clarity,
we first discuss in this section spherical fireballs of energy $E$
expanding into uniform ambient density $n$. We then generalize the discussion
to the case where spherical symmetry and homogeneity are not assumed.
Since the value of $p$ in the cases where it is best determined by
observations is $p=2.2$, we present numeric results for $p=2.2$ and
comment on the sensitivity of the results to changes in $p$ value.

Adopting the hypothesis that the electron spectral index value is universal,
$p\approx2$, the fact that $\beta_{OX}>0.6$ in all cases (see Table 1)
implies that $\nu_c<\nu_X$ in all cases (Note that extinction in
the host galaxy can only reduce the observed value of $\beta_{OX}$).
This is indeed expected, since
on time scale of a day, the time scale over which X-ray afterglow is observed,
$\nu_c$ is typically expected to be well bellow the X-ray band.
The (fireball rest frame) Lorentz factor
$\gamma_c$ of electrons for which the synchrotron cooling time is comparable
to the (rest frame) adiabatic cooling time is determined
by $6\pi m_e c/\sigma_T\gamma_cB^2=24\Gamma t/13$.
Here, $\Gamma$ is the shocked plasma Lorentz factor, related to
observed time by $t\approx r/4\Gamma^2c$ (\cite{Waxman97c}), and
the adiabatic cooling time is $6r/13\Gamma c$ (\cite{GnW99}).
The characteristic (observed) frequency of synchrotron
photons emitted by such electrons,
$\nu_c\approx0.3\Gamma\gamma_c^2eB/2\pi m_ec$, is
\begin{equation}
\nu_c\approx4.7\times10^{13}\left({1+z\over2.5}\right)^{-1/2}
\xi_{B,-2}^{-3/2}n_0^{-1}E_{53}^{-1/2}
t_{\rm d}^{-1/2}\quad{\rm Hz},
\label{eq:nu_c}
\end{equation}
where $E=10^{53}E_{53}$~erg, $n=1n_0{\rm cm}^{-3}$, $\xi_B=10^{-2}\xi_{B,-2}$
and $t=1t_{\rm d}$~d. In deriving Eq. (\ref{eq:nu_c})
we have used the self-similar relation between fireball Lorentz factor
and radius, $\Gamma=(17E/16\pi n m_p c^2)^{1/2}r^{-3/2}$ (\cite{BnM76}),
and the relation
$t=r/4\Gamma^2c$ between fireball radius and observed time (\cite{Waxman97c}).

The synchrotron peak flux $f_m$
in the fireball model under consideration is time independent,
\begin{equation}
f_m=C_1(1+z)d_L^{-2}\xi_B^{1/2}En^{1/2},
\label{eq:f_m}
\end{equation}
and the peak frequency $\nu_m$ decreases with time,
\begin{equation}
\nu_m=C_2(1+z)^{1/2}\xi_e^2\xi_B^{1/2}E^{1/2}t^{-3/2}.
\label{eq:nu_m}
\end{equation}
Here, $C_{1,2}$ are numeric constants, and $d_L$ is the burst luminosity
distance.
In what follows we use the analytic results of Gruzinov \& Waxman
(1999)
for the values of the numeric constants, $C_1=1.4\times10^{-21}{\rm cm}^{3/2}$
and $C_2=6.1\times10^{-5}{\rm s^{3/2} g^{-1/2} cm^{-1}}$. Similar
values have been obtained by numerical (\cite{GPS99}) and approximate
analytic (\cite{WnG99}) calculations. Order unity differences between the
various analyses reflect different detailed model assumptions and degree
of accuracy of the approximations. Note, that the exact value
of $C_2$ depends on the detailed shape of the electron distribution
function ($C_1$ is insensitive to such details, \cite{GnW99}).
Although we have strong evidence that at high energy this
distribution is a power-law, it may not be a pure power-law at low energy,
which would affect the value of $C_2$. However, the relation (\ref{eq:nu_m})
would still hold for all GRBs as long as the distribution function is
universal.

Using Eqs. (\ref{eq:nu_c}--\ref{eq:nu_m}), the fireball electron energy,
$\xi_eE$, is related to the observed
flux $f_\nu(\nu,t)$ at time $t$ and frequency $\nu\gg\nu_c$,
$f_\nu=f_m\nu_m^{(p-1)/2}\nu_c^{1/2}\nu^{-p/2}$, as
\begin{equation}
\xi_eE=(C_2C_3)^{-1/2}C_1^{-1}{d_L^2\over1+z}\nu t f_\nu(\nu,t)
Y^\epsilon,
\label{eq:E_e}
\end{equation}
where
\begin{equation}
Y\equiv C_1C_3^{1/2}C_2^{-3/2}\xi_e^{-3}\xi_B^{-1}
d_L^{-2}\nu t^2 f^{-1}_\nu(\nu,t)\,,\quad \epsilon\equiv{\frac{p-2}{p+2}}.
\label{eq:Y}
\end{equation}
Here, we have defined $C_3=6.9\times10^{39}{\rm s^{-3/2} g^{1/2} cm^{-2}}$
so that
$\nu_c=C_3(1+z)^{-1/2}\xi_B^{-3/2}n^{-1}E^{-1/2}t^{-1/2}$.
Eq. (\ref{eq:E_e}) implies that a measurement of the flux $f_\nu$ at
a frequency above the cooling frequency provides a robust estimate of the
fireball electron energy. The energy estimate is independent of the
ambient density $n$ and nearly independent of $\xi_B$,
$\xi_eE\propto\xi_B^\epsilon$ with $\epsilon\ll1$ (e.g. $\epsilon=1/21$ for
p=2.2). Since $\epsilon\ll1$, the value of $Y^\epsilon$ is similar for
all GRBs. It also implies that changing the value of $p$
would affect the energy estimate of all bursts in a similar way.
For typical parameters, $f(\nu=10^{18}{\rm Hz},t=1{\rm d})=0.1\mu{\rm Jy}$,
$d_L=3\times10^{28}$~cm, $\xi_e=0.2$ and $\xi_B=0.01$, we have
$Y^\epsilon=10^{10\epsilon}$, i.e. $Y^\epsilon=3$ for $p=2.2$ and
$Y^\epsilon=8$ for $p=2.4$. Note that changing $p=2.2$ to $p=2.4$ would
increase the energy estimate by less than a factor of $8/3$, since the
value of $C_2$ is higher for larger $p$. For pure power-law electron
distribution, $C_2^{1/2}\propto(p-2)/(p-1)$ and the energy obtained assuming
$p=2.4$ is larger than that obtained assuming $p=2.2$ by a factor of $1.6$.

A few comments should be made here regarding the uniform density
and spherical symmetry assumptions.
Since the energy estimate (\ref{eq:E_e}) is independent of $n$, it holds
not only for the case of homogeneous density, but also to models with
variable density, e.g. for wind models in which $n\propto r^{-2}$. The
values of the constants $C_{1,2,3}$ would of course differ, by order unity
factors, from those used here. However, the relation (\ref{eq:E_e}) should
still hold and the energy estimate would not be significantly affected
as $C_{1,2,3}$ would not be significantly modified.

We have so far assumed spherical symmetry. Since the fireball expands at
relativistic speed, a distant observer receives radiation from a conical
section of the fireball with opening angle $\sim1/\Gamma(t)$ around the line
of sight. Thus, the energy $E$ in the discussion above should be understood
as the energy that the fireball would have carried had it been spherically
symmetric. In particular, Eq. (\ref{eq:E_e}) determines
the fireball energy per unit solid angle
$\varepsilon_e\equiv\xi_eE/4\pi$,
within the observable cone of opening
angle $1/\Gamma(t)$. The Lorentz factor $\Gamma$,
\begin{equation}
\Gamma=10.6\left({1+z\over2}\right)^{3/8}
\left({E_{53}\over n_0}\right)^{1/8}t_{\rm d}^{-3/8},
\label{eq:Gamma}
\end{equation}
is only weakly dependent on fireball parameters. Thus, X-ray observations
on $\sim1$~d time scale provide information on a conical section of the
fireball of opening angle $\theta\sim0.1$ (this holds also for the wind case,
\cite{LnW99}). Note, that since observed $\gamma$-rays are emitted at
a stage the fireball is highly relativistic, $\Gamma\sim300$, $\gamma$-ray
observations provide information on a much smaller section of the
fireball, $\theta\sim10^{-2.5}$.

Table 2 compares observed GRB $\gamma$-ray energy per unit
solid angle with fireball electron
energy per unit solid angle
derived from X-ray afterglow flux using Eq. (\ref{eq:E_e}).
Results are given for all bursts with published $\gamma$-ray fluence and
X-ray afterglow flux (Note, that for all but one of the BeppoSAX triggered GRBs
which were observed with the NFI, X-Ray afterglow has been detected;
We exclude GRB~980425 from the analysis,
since for this bursts it is not clear whether or not X-ray afterglow
was detected, \cite{Pian}).
We present our results in terms of
$3\varepsilon_e=3\xi_e\varepsilon$, which is
the total fireball energy under the assumption of
electron energy equipartition, $\xi_e\sim1/3$.
The results are also shown in Fig. 1. We have included in the table and
figure bursts for which optical data is not available. For such bursts,
the effective spectral index $\beta_{OX}$ can not be determined, and therefore
it can not be directly demonstrated from observations that $\nu_X>\nu_c$,
and hence that Eq. (\ref{eq:E_e}) applies. However, it is clear from
Eq. (\ref{eq:nu_c}) that the condition $\nu_X>\nu_c$ is likely to be
satisfied on a day time scale. In addition, the distribution of GRB
$\gamma$-ray and total energy ratios
we infer for bursts without optical counterpart is similar to that of
bursts with optical counterpart, indicating that indeed the condition
$\nu_X>\nu_c$ is satisfied.
To determine
the absolute energy of GRBs for which the redshift is unknown, we have assumed
GRB redshift of $z=1.5$, since, based on measured GRB redshifts,
most detected GRBs are expected to occur at the redshift range of 1 to 2
(\cite{KTH98,MnM98,HnF99}).

\subsection{Electron energy fraction}

Several characteristics of afterglow observations imply that the
electron energy fraction $\xi_e$ is close to equipartition.
For GRB~970508 afterglow data is detailed enough
to determine $\xi_e\sim0.2$ (\cite{Waxman97b,WnG99,GPS99}). A similar
conclusion can be drawn for GRB~971214. For this GRB,
$\nu_m\approx4\times10^{14}$~Hz
and $f_m\approx0.03$~mJy have been observed at
$t=0.58$~d (\cite{Ramaprakash}). Using Eqs. (\ref{eq:f_m})
and (\ref{eq:nu_m}) this implies, for
GRB~971214 redshift $z=3.42$ (\cite{Kulkarni971214}),
$\xi_e\sim1(\xi_B/0.1)^{-1/8}n_0^{1/8}$. Thus,
a value close to equipartition is implied by GRB~971214 observations
(\cite{WnG99} suggest a different interpretation of the GRB~971214 data,
which also requires $\xi_e\sim1$).

For most other bursts, afterglow observations are not sufficient for
determining the value of $\xi_e$. However, $\xi_e\sim1/3$ is consistent with
all afterglow data, and is indeed commonly used in afterglow models
attempting to account for observations. This is consistent with our hypothesis
of universal $\xi_e$ value close to equipartition. Additional support for
this hypothesis is provided by the following argument.

GRANAT/SIGMA observations of GRB~920723 (\cite{Granat99}) and BATSE
observation of GRB~980923 (\cite{Giblin99}) show a continuous transition
from GRB to afterglow phase in the hard X-ray range. It has also been shown
for several BeppoSAX bursts that the 2--10~keV flux at the end of the
GRB phase, on time scale of tens of seconds,
is close to that obtained by extrapolating
backward in time the X-ray afterglow flux
(e.g. \cite{Costa97,Zand980329}).
This suggests that the late GRB flux is in fact
dominated by afterglow emission, i.e. by synchrotron emission from
the shock driven into the ambient medium (see also \cite{Frontera99}).
On minute time scale, the emission peaks at $\sim10$~keV energies
(\cite{Giblin99,Frontera99}), implying $\nu_m\sim2.5\times10^{18}$~Hz
at this time (note that this
value is consistent with the $\nu_m$ values inferred at later time
for GRB~970508 and GRB~971214). Using Eq. (\ref{eq:nu_m}), we find that
this in turn implies $\xi_e\sim0.4(\xi_B/0.1)^{-1/4}E_{53}^{-1/4}$. Thus,
a value of $\xi_e$ well below equipartition, $3\xi_e\ll1$,
would require $E_{53}\propto\xi_e^{-4}\gg1$, inconsistent with
our conclusion that $3\xi_eE_{53}\sim1$ (see Fig. 1).


\section{Discussion and implications}

We have shown that afterglow observations strongly suggest that the energy
distribution of shock accelerated electrons is universal, given at high energy
by a power-law
$dn_e/d\gamma_e\propto\gamma_e^{-p}$ with $p\approx2.2$ (\S2.1), and
that the energy fraction carried by electrons is also universal and
close to equipartition, $\xi_e\sim1/3$ (\S2.3). Adopting the hypothesis that
the value of $p$ is universal and close to 2, $p\approx2.2$,
we showed (\S2.2) that a single measurement at time $t$
of X-ray afterglow
flux provides, through Eq. (\ref{eq:E_e}),
a robust estimate of the fireball electron energy per
unit solid angle, $\varepsilon_e$,
over a conical section of the fireball of opening angle
$\sim1/\Gamma(t)$, where $\Gamma(t)$ is the fireball Lorentz factor.
On day time scale $\Gamma\sim10$ [see Eq. (\ref{eq:Gamma})], and X-ray flux
therefore provides a robust estimate of $\varepsilon_e$ over an opening
angle $\theta\sim0.1$. Adopting the hypothesis that $\xi_e$ is close
to equipartition, the X-ray flux provides also a robust estimate of the
total fireball energy per unit solid angle, $\varepsilon$, over an
opening angle $\theta\sim0.1$.

We emphasize here that the total (or electron) fireball energy estimates
are not based on the total X-ray fluence. The X-ray fluence is dominated
by the early time, $t\sim10$~s, emission. Since at this time the fireball
Lorentz factor is high, $\Gamma\sim10^{2.5}$, the total X-ray fluence provides
information only on a small section of the fireball,
$\theta\sim10^{-2.5}\ll0.1$. Moreover, using the X-ray fluence to derive
constraints on fireball parameters using afterglow models is complicated since
on $\sim10$~s time scale the fireball is not in the self-similar expansion
stage for which afterglow models apply, since GRB emission on this time scale
is important and it is difficult to separate afterglow and main GRB
contributions, and since the X-ray afterglow flux is not observed in the
time interval of $\sim10^2$~s to $\sim0.5$~d, which implies that
the total fluence depends strongly on the interpolation of X-ray flux
over the time interval in which it is not measured.

It is also important to emphasize that while the X-ray flux is independent
of the ambient density into which the fireball expands and very
weakly dependent
on the magnetic field energy fraction [see Eq. (\ref{eq:E_e})], the optical
flux detected on a time scale of a day is sensitive to both parameters,
since the cooling frequency is close to the optical band on this time scale
[see Eq. (\ref{eq:nu_c})]. Moreover, the optical flux may be significantly
affected by extinction in the host galaxy. Thus, while the observed X-ray
flux provides mainly information on intrinsic fireball parameters,
the optical flux depends strongly on the fireball environment.

Our results for $\varepsilon$ are presented in Table 2 and in Fig. 1,
where $\gamma$-ray
energy per unit solid angle, $\varepsilon_\gamma$, is plotted as a function
of fireball energy per unit solid angle, $\varepsilon$. Several conclusion
may be drawn based on the table and figure:
\begin{enumerate}

\item Fireball energies of observed GRBs are in the range of
$4\pi\varepsilon=10^{51.5}$ to $10^{53.5}$~erg.

\item $\varepsilon_\gamma$ and $\varepsilon$ are strongly correlated, with
$|\log_{10}(\varepsilon_\gamma/\varepsilon)|\lesssim0.5$. Thus,
the most energetic bursts
(i.e. those with highest isotropic $\gamma$-ray energy)
represent much higher energy release from the sources.

\item Our results are inconsistent with
models in which GRB $\gamma$-ray emission is produced by internal shocks
with low efficiency (\cite{Spada,Kumar,KnP99}),
$\varepsilon_\gamma/\varepsilon<10^{-2}\ll1$  for electron
energy equipartition $\xi_e\sim1/3$ (and still lower efficiency for lower
electron energy fraction).
However, we believe this
contradiction should not be considered strong evidence against internal
shock models, since the low efficiency obtained in the analyses
reflects mainly the underlying assumptions of the models regarding the
variability of the wind produced by the source. In particular, if the
typical ratio of Lorentz factors of adjacent fireball wind
shells is large,
rather than being of order unity as commonly assumed, the internal shock
efficiency may be close to unity.

\item The strong correlation between $\varepsilon_\gamma$ and $\varepsilon$,
and in particular the fact that values $\varepsilon_\gamma/\varepsilon\gg1$
are not obtained, implies that if fireballs are jets of
finite opening angle $\theta$, then $\theta$ should satisfy
$\theta\gtrsim0.1$. This is due to the fact that
for $\theta\ll0.1$ significant sideways expansion of the jet
would occur well before X-ray afterglow observations, during which
$\Gamma\sim10$, leading to
$\varepsilon_\gamma/\varepsilon\gg1$. Our conclusion is similar to
that obtained by Piro {\it et al.} (1999) based on a different argument
(Piro {\it et al.} rely on analysis of the relation between
X-ray afterglow spectral and temporal characteristics).

\item The fact that $\varepsilon_\gamma/\varepsilon$ values larger than
unity, $\varepsilon_\gamma/\varepsilon\sim3$, are obtained for
$4\pi\varepsilon_\gamma>10^{52}$~erg, may suggest that the electron
energy fraction in these cases is somewhat below equipartition
($3\xi_e=1/3$ can not be ruled out by our analysis),
and/or that the fireball
is a jet of opening angle $\theta_j\sim0.1$, in which case some
reduction of energy per solid angle at the X-ray observation time would be
detectable. Note, that GRB~990123 and GRB~980519, for which
evidence for a jet-like fireball exists and which are included in our
analysis, are indeed in the group for which
$\varepsilon_\gamma/\varepsilon\sim3$.

\item For low energy,
$4\pi\varepsilon_\gamma<10^{52}$~erg, bursts we find
$\varepsilon_\gamma/\varepsilon\sim1/3$. This suggests that
the $\gamma$-ray production efficiency of such bursts is lower than that
of higher energy bursts. We note that, based on published GRB light curves,
there is evidence that the lower energy,
$4\pi\varepsilon_\gamma<10^{52}$~erg, bursts are characterized by shorter
durations of $\gamma$-ray emission, compared to that of
$4\pi\varepsilon_\gamma>10^{52}$~erg bursts. A more detailed analysis of
light curves of all GRBs in our sample should be carried out to
confirm or rule out such correlation.

\item While ``low efficiency,'' $\varepsilon_\gamma/\varepsilon\ll1$,
bursts do not exist for fireball energy $4\pi\varepsilon>10^{52.5}$~erg,
$\varepsilon_\gamma/\varepsilon\ll1$
bursts (as suggested by \cite{Spada,Kumar,KnP99})
may exist for lower fireball energy,
$4\pi\varepsilon<10^{52.5}$~erg, as they would
not have been detected by BeppoSAX (as indicated in Fig. 1).

\end{enumerate}

Our analysis allows to determine the total fireball energy per unit solid
angle averaged over an angle $\theta\sim0.1$. A determination of the total
energy emitted from the source requires knowledge of the fireball opening
angle. If typical opening angles are $\theta\sim0.1$, an angle suggested by
the observed breaks in optical light curves and consistent with our
results, the total energy emitted by the underlying
GRB sources is in the range of $10^{50}$~erg to $10^{51.5}$~erg.
The total energy may be two
orders of magnitude higher, however, if emission from the source is close
to spherical. X-ray observations on time scale $\gg1$~d, during which
$\Gamma\ll10$, will provide information on fireball properties averaged over
larger opening angles and will therefore provide
stronger constraints on the total energy associated with GRB explosions.

\acknowledgements
We thank J. N. Bahcall, D. A. Frail, A. Gruzinov, A. Loeb and B. Paczy\'nski
for useful comments on a previous version of this manuscript.
DLF acknowledges support from the Karyn Kupcinet
fund during her stay at the Weizmann Institute. EW is partially supported
by BSF Grant 9800343, AEC Grant 38/99 and MINERVA Grant.

\newpage

\begin{table}[ht]
\begin{center}
\begin{tabular}{l|llllll}  \hline \hline
\multicolumn{1}{c|}{{\bf GRB}} & \multicolumn{1}{c}{{\bf X-ray}}
& \multicolumn{1}{c}{{\bf X-ray\tablenotemark{b}}}
& \multicolumn{1}{c}{{\bf Optical}} & \multicolumn{1}{c}{{\bf Optical}}
& \multicolumn{1}{c}{{\bf Interpolated\tablenotemark{d}}}
& \multicolumn{1}{c}{{\bf Spectral}}  \\
& \multicolumn{1}{c}{{\bf Time\tablenotemark{a}}} &  \multicolumn{1}{c}{{\bf Flux}}
& \multicolumn{1}{c}{{\bf Time}} & \multicolumn{1}{c}{{\bf Flux\tablenotemark{c}}}
& \multicolumn{1}{c}{{\bf X-ray Flux}} &  \multicolumn{1}{c}{{\bf Index\tablenotemark{e}}} \\
 & & \multicolumn{1}{c}{{\bf ($10^{-12}$ergs}} & & &  \multicolumn{1}{c}{{\bf ($10^{-12}$ergs}} & \\
 & \multicolumn{1}{c}{{\bf (days)}} & \multicolumn{1}{c}{{\bf cm$^{-2}$ s$^{-1}$)}}
& \multicolumn{1}{c}{{\bf (days)}} & \multicolumn{1}{c}{{\bf ($\mu$Jy)}}
& \multicolumn{1}{c}{{\bf cm$^{-2}$s$^{-1}$)}} & \multicolumn{1}{c}{$\beta_{OX}$}  \\   \hline
\textbf{970228} & 0.50  & $2.8 \pm0.4^1$    	& 0.69  & $26 \pm50$\% (R$_{c}$)$^{14}$ & 1.82 & 0.64 \\
\textbf{970508} & 6.1   & 0.0065$^2$ 		& 6.08  & $9.6 \pm5$\% (R$_{c}$)$^{14}$  & 0.0066 & 1.2 \\
\textbf{970616} & 0.17  & 11$^3$          	& * 	& * 	  &  *  & * \\
\textbf{970828} & 0.16  & 5.3$^4$ 		& * 	& * 	  &   * & * \\
\textbf{971214} & 0.44  & 0.7$^5$  		& 0.54  & $4.6 \pm6$\% (R)$^{15}$  & 0.54   & 0.58  \\
\textbf{971227} & 0.67  & 0.26$^6$  		& * 	& *    	  &  *  & * \\
\textbf{980329} & 0.29  & $1.4 \pm0.2^7$    	& 0.71  & $12 \pm30$\% (I)$^{16}$ & 0.34   & 0.74  \\
\textbf{980519} & 0.48  & $0.38 \pm0.06^8$  	& 0.49  & $97 \pm3$\% (I)$^{17}$  & 0.37   & 0.99  \\
\textbf{980613} & 0.41  & 0.33$^9$ 	    	& 1.0   & $3.7 \pm9$\% (R)$^{18}$  & 0.10   & 0.77  \\
\textbf{980703} & 1.15  & $0.47 \pm0.07^{10}$   & 1.2   & $16 \pm3$\% (I)$^{19}$  & 0.45    & 0.74 \\
\textbf{990123} & 0.25  & 11$^{11}$   	 	& 0.55  & $32 \pm5$\% (gunn r)$^{20}$ & 3.8   & 0.58 \\
\textbf{990506} & 0.09  & $35.1 \pm15^{\rm f,12}$ & * 	&  *   &   * & * \\
\textbf{990806} & 0.32  & $0.55 \pm0.15^{\rm g,13}$ & * & *    &  *  &  *\\
\hline
\end{tabular}
\end{center}
\caption{X-ray and optical afterglow data.}
\label{spectral}
\tablenotetext{*} {Data not available.}
\tablenotetext{a} {Since X-ray flux varies significantly over typical
 integration
 time, the quoted time is an ``effective'' time, the time at which the
 instantaneous flux equals the average (over the integration time) flux,
 given in the second column. The effective time is calculated assuming power-
 law decay in time, using the index reported based on X-ray observations.}
\tablenotetext{b} {2--10~keV, unless otherwise noted.}
\tablenotetext{c} {Corrected for Galactic extinction.  Extinction in the GRB
 host galaxy may reduce optical flux and result in lower observed
 $\beta_{OX}$. To minimize this effect, we use longest wavelength for which
 accurate data is available.}
\tablenotetext{d} {X-ray flux interpolated to optical measurement time.}
\tablenotetext{e} {Error bars not available for many X-ray detections. A
 20\% error in flux ratio corresponds to 0.02 error in $\beta_{OX}$,
 while a factor of 2 error in flux ratio corresponds to 0.09
 error in $\beta_{OX}$.}
\tablenotetext{f} {2--12~keV.}
\tablenotetext{g} {1.4--10~keV.}
\tablenotetext{}{
$^{1}${\cite{Costa97}};
$^{2}${\cite{table6}};
$^{3}${\cite{table62}};
$^{4}${\cite{table26}};
$^{5}${\cite{table131}};
$^{6}${\cite{table132}};
$^{7}${\cite{Zand980329}};
$^{8}${\cite{table58}};
$^{9}${\cite{table28}};
$^{10}${\cite{newtableVrees}};
$^{11}${\cite{table47}};
$^{12}${\cite{newtable71}};
$^{13}${\cite{table102}};
$^{14}${\cite{table42}};
$^{15}${\cite{newtable10}};
$^{16}${\cite{table18}};
$^{17}${\cite{Halpern99}};
$^{18}${\cite{table33}};
$^{19}${\cite{newtableVrees}};
$^{20}${\cite{table140}}.
}
\end{table}

\newpage
\begin{table}[h]
\begin{center}
\begin{tabular}{l|lllll}  \hline \hline
\multicolumn{1}{c|}{{\bf GRB}} & \multicolumn{1}{c}{{\bf $\gamma$-Ray\tablenotemark{a} Fluence}}
& \multicolumn{1}{c}{{\bf Redshift}} & \multicolumn{1}{c}{{\bf $\gamma$-ray Energy}}
& \multicolumn{1}{c}{{\bf Fireball Energy}} & \\
& \multicolumn{1}{c}{{\bf ($10^{-6}$ ergs}} & \multicolumn{1}{c}{$z$}
& \multicolumn{1}{c}{{\bf $4\pi\varepsilon_\gamma$}}
& \multicolumn{1}{c}{{\bf $4\pi\varepsilon$}}
& \multicolumn{1}{c}{{\bf $\varepsilon_\gamma/\varepsilon$}}  \\
& \multicolumn{1}{c}{{\bf cm$^{-2}$ s$^{-1}$)}} &
& \multicolumn{1}{c}{{\bf ($10^{52}$ ergs)}}
& \multicolumn{1}{c}{{\bf ($10^{52}$ ergs)}} &   \\  \hline
\textbf{970228} & 11$^{\rm b, 1}$  		& 0.695$^{14}$ 	& 1.3 			& $4.2 \pm0.6$ 		& $0.31\pm0.05$  \\
\textbf{970508} & $1.8 \pm0.3^{\rm b, 2}$   & 0.835$^{15}$  & $0.30 \pm0.049$ 	& $0.62 \pm0.04^{\rm c}$ 	& $0.48\pm0.14$ \\
\textbf{970616} & $40.1 \pm1.4^3$	& * 		& $11 \pm0.39$* 	& 16* 			& 0.67 \\
\textbf{970828} & 70$^4$ 		& 0.958$^{19}$ 		& 8.9 			& 3.8* 			& 2.3 \\
\textbf{971214} & $10.9 \pm0.7^5$  	& 3.42$^{16}$   & $11 \pm0.69$          & 11 			& 0.99 \\
\textbf{971227} & $0.93 \pm0.14^6$   	& * 		& $0.27 \pm0.39$* 	& 2.1* 			& 0.13 \\
\textbf{980329} & $55 \pm5^{\rm b, 7}$  & * 		& $24 \pm2.2$* 		& $4.1 \pm0.6$* 	& $5.8\pm1.2$ \\
\textbf{980519} & $25.4 \pm4.1^8$       & * 		& $7.3 \pm1.2$* 	& $2.1 \pm0.3$* 	& $3.5\pm1.2$   \\
\textbf{980613} & $1.71 \pm0.25^{9}$ 	& 1.096$^{15}$  & $0.28 \pm0.041$  	& 0.93 			& 0.30  \\
\textbf{980703} & $45.9 \pm0.42^{10}$  	& 0.966$^{17}$ 	& $6.0 \pm0.055$ 	& 3.2 			& 1.87  \\
\textbf{990123} & $300^{11}$  		& 1.61$^{18}$ 	& $90$ 	& 28			& 3.3 \\
\textbf{990506} & $223 \pm2^{12}$  	& * 		& $61 \pm0.55$* 	& $28 \pm9$* 		& 2.5 \\
\textbf{990806} & $3.29 \pm0.53^{13}$   & * 		& $0.90 \pm0.14$* 	& $1.9 \pm0.4$* 	& 0.47  \\
\hline
\end{tabular}
\end{center}
\caption{$\gamma$-ray and fireball energy per unit solid angle.
 Energies calculated for flat universe, zero cosmological constant,
 $H_0=75{\rm km/s\,Mpc}$. $p=2.2$ assumed for fireball energy estimated from X-ray flux.}    \label{energy}
\tablenotetext{*} {Redshift unknown, $z$ = 1.5 assumed for energy
 calculations.}
\tablenotetext{a}{20~keV to 2~MeV unless otherwise noted.}
\tablenotetext{b}{40--700~keV; extrapolated in energy calculatuion
 to 20~keV--2~MeV range assuming power-law photon spectrum of index 2.}
\tablenotetext{c}{X-ray flux used in this calculation is: $0.63\pm0.06$ at
 0.24 day at 2--10~keV (\cite{table129})}.
\tablenotetext{}{
$^{1}${\cite{Costa97}};
$^{2}${\cite{table6}};
$^{3}${\cite{table61}};
$^{4}${\cite{table25}};
$^{5}${\cite{Ramaprakash}};
$^{6}${\cite{table66}};
$^{7}${\cite{Zand980329}};
$^{8}${\cite{Halpern99}};
$^{9}${\cite{table31}};
$^{10}${\cite{table44}};
$^{11}${\cite{new26Briggs}};
$^{12}${\cite{table55}};
$^{13}${\cite{table101}};
$^{14}${\cite{newtable25}};
$^{15}${\cite{newtable22i}};
$^{16}${\cite{Ramaprakash}};
$^{17}${\cite{newtableVrees}};
$^{18}${\cite{newtable2}};
$^{19}${\cite{Djorgovski2000}}.
}
\end{table}

\newpage

\begin{figure}
\centerline{\psfig{figure=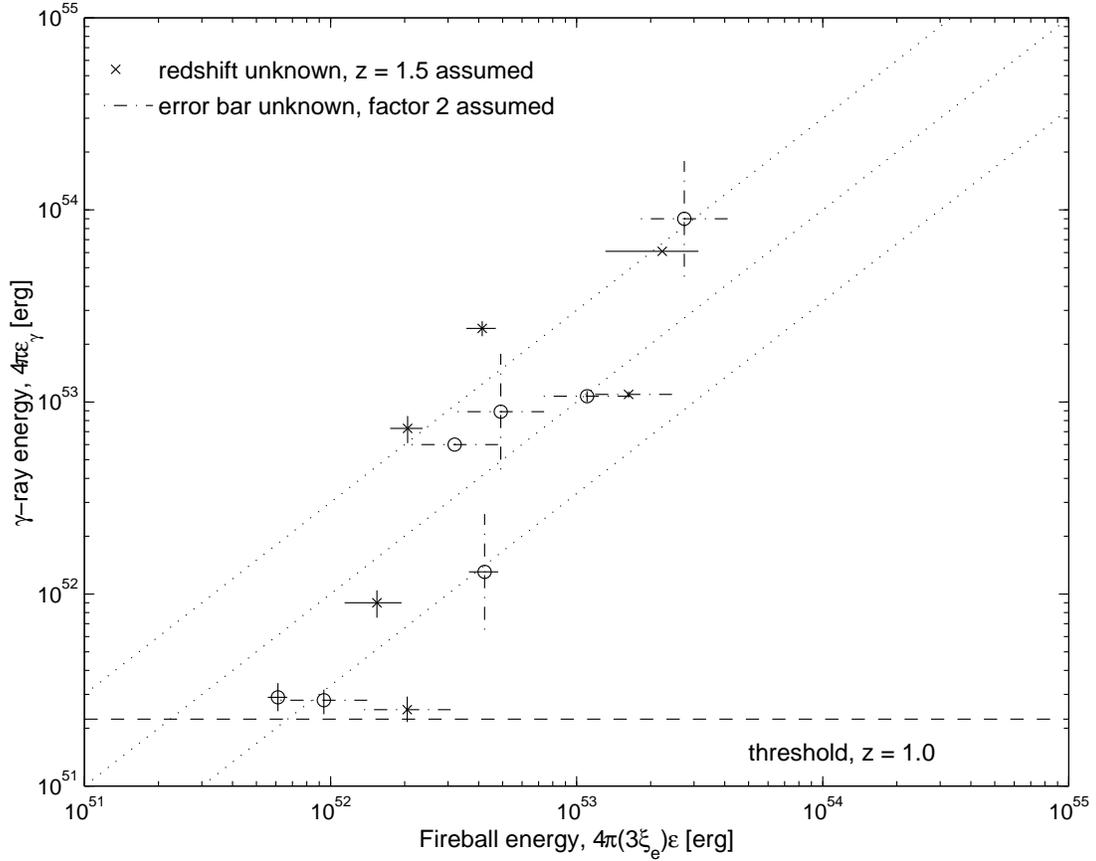,width=6in}}
\caption{
The relation between observed GRB $\gamma$-ray energy per unit solid angle,
$\varepsilon_\gamma$, and fireball energy per unit solid angle,
$\varepsilon$, inferred from X-ray flux measurement at delay $\sim1$~d.
Energies calculated for flat universe, zero cosmological constant,
$H_0=75{\rm km/s\,Mpc}$, and assuming $p=2.2$ [$\varepsilon$ is increased by
$\approx50\%$ for $p=2.4$, see discussion following Eq. (\ref{eq:E_e})].
BeppoSAX detection threshold is shown for burst duration of 10~s.
Dotted lines correspond to $\varepsilon_\gamma/\varepsilon=1/3$, 1 and 3.
}
\label{fig1}
\end{figure}

\end{document}